\documentclass[11pt]{article}  
\usepackage{menuproc}
\usepackage{cite}
\usepackage{epsfig}
\usepackage{amsmath,amssymb}
\begin{document}
%
%
%
\titlematter{The $\chi$-BS(3) approach}%
{M.F.M. Lutz$^a$ and E.E. Kolomeitsev$^b$ }%
{$^a$ Gesellschaft f\"ur Schwerionenforschung (GSI),
Planck Str. 1, D-64291 Darmstadt, Germany\\
 $^b$ECT$^*$, Villa Tambosi, I-38050 \,Villazzano  (Trento)
and INFN, G.C.\ Trento, Italy}%
{
We present the results of the $\chi$-BS(3) approach demonstrating that
 a combined chiral and $1/N_c$ expansion of the Bethe-Salpeter
interaction kernel leads to a good description of the kaon-nucleon, antikaon-nucleon and pion-nucleon
scattering data typically up to laboratory momenta of $p_{\rm lab} \simeq $ 500 MeV.
We solve the covariant on-shell reduced coupled channel Bethe-Salpeter equation with the interaction kernel
truncated to chiral order $Q^3$ and to the leading order
in the $1/N_c$ expansion}
%
%


The main features and crucial
arguments of our recent work on meson-baryon scattering \cite{LuKol} are briefly summarized.
Within the $\chi$-BS(3) approach we consider the
number of colors ($N_c$) in QCD as a large
parameter relying on a systematic expansion of the interaction kernel in powers of
$1/N_c$. The coupled-channel Bethe-Salpeter kernel is evaluated in a combined chiral
and $1/N_c$ expansion including terms of chiral order $Q^3$.

We expect
all baryon resonances, with the important exception of those resonances which belong
to the large $N_c$ baryon ground states, to be generated by coupled channel dynamics.
This conjecture is based on the observation that unitary (reducible) loop diagrams are
typically enhanced by a factor of $2 \pi $ close to threshold relatively to
irreducible diagrams. That factor invalidates the perturbative evaluation  of the
scattering amplitudes and leads necessarily to a non-perturbative scheme with
reducible diagrams summed to all orders. In our present scheme
we consider an explicit s-channel baryon nonet term with $J^P={\textstyle{3\over 2}}^-$
in the interaction kernel as a reminiscence of further inelastic channels not included
like for example the $K\,\Delta_\mu $ or $K_\mu \,N$ channel.

The scattering amplitudes for the meson-baryon scattering processes are obtained from
the solution of the coupled channel Bethe-Salpeter scattering equation. Approximate
crossing symmetry of the amplitudes is guaranteed by a renormalization program which
leads to the matching of subthreshold amplitudes. A further important ingredient of
our scheme is a systematic and covariant on-shell reduction of the Bethe-Salpeter
equation. We point out that an on-shell reduction is mandatory as to avoid an
unphysical and uncontrolled dependence on the choice of chiral coordinates or the
choice of interpolating fields. In other words given our scheme the on-shell
scattering amplitude will not change if we used a different representation of the
chiral Lagrangian. In the $\chi$-BS(3) scheme the on-shell reduction is implied
unambiguously by the existence of a unique and covariant projector algebra which
solves the Bethe-Salpeter equation for any choice of quasi-local interaction terms.

At subleading order $Q^2$ the chiral $SU(3)$
Lagrangian predicts the relevance of 12 basically unknown parameters,
which all need to be adjusted to the empirical scattering data.
It is important to realize that chiral symmetry is largely predictive in the $SU(3)$ sector
in the  sense that it reduces the number of parameters  beyond
the static $SU(3)$ symmetry. For example one should compare the six tensors which
result from decomposing $8\otimes 8= 1
\oplus 8_S\oplus 8_A \oplus 10\oplus \overline{10}\oplus 27$ into its
irreducible components with the subset of SU(3) structures selected
by chiral symmetry in a given partial wave. Thus static $SU(3)$
symmetry alone would predict 18 independent terms for the s-wave
and two p-wave channels rather than the 12 chiral $Q^2$ background
parameters. In our work the number of parameters was
further reduced significantly by insisting on the large $N_c$ sum rules
for the symmetry conserving quasi-local two body interaction terms
leaving only 5 parameters.  All parameters are found to have natural size.

\begin{figure}[t]
\parbox{.5\textwidth}{\epsfig{file= 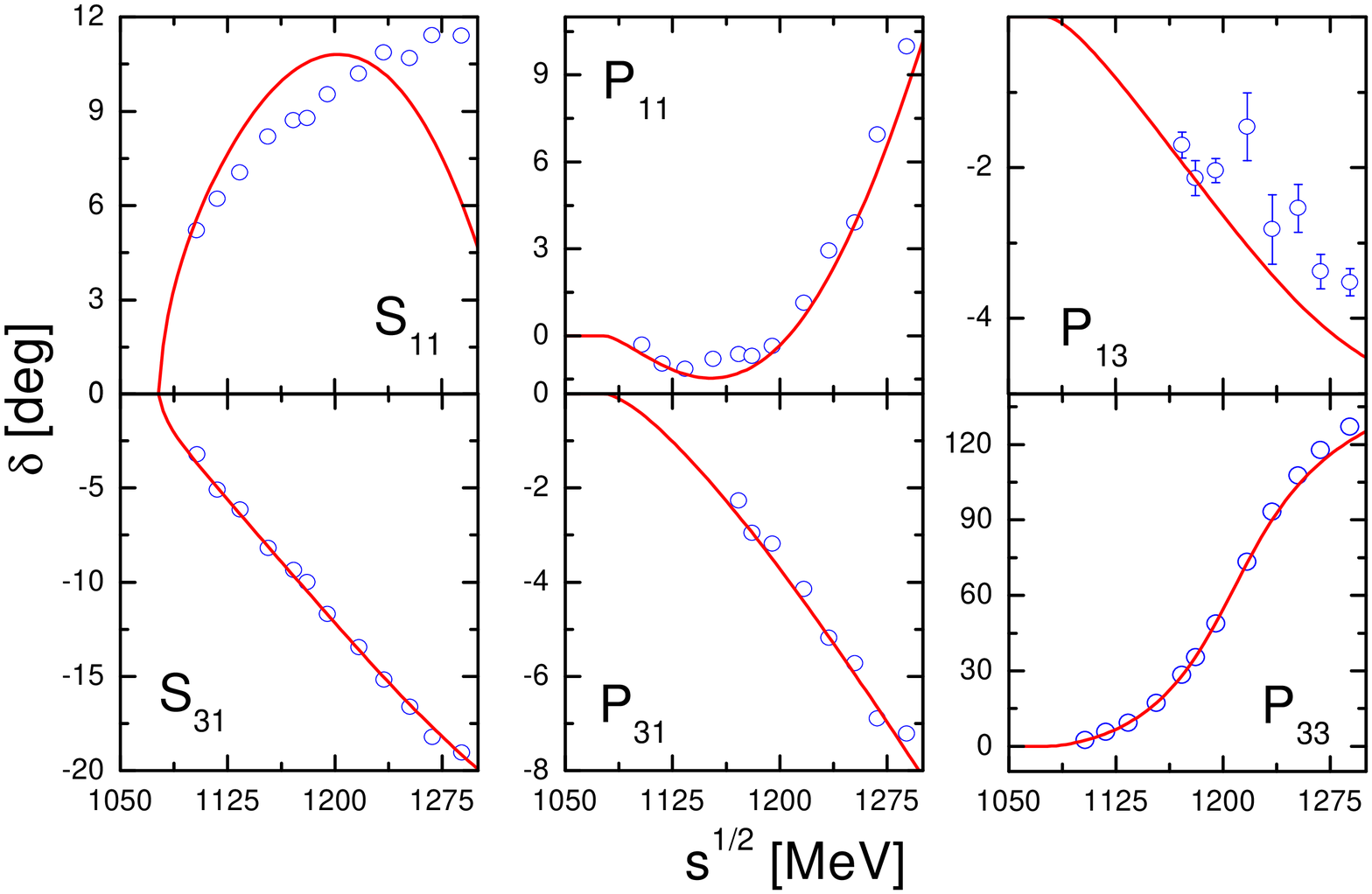,width=.48\textwidth,silent=,clip=}}
\hfill
\parbox{.5\textwidth}{
\epsfig{file= 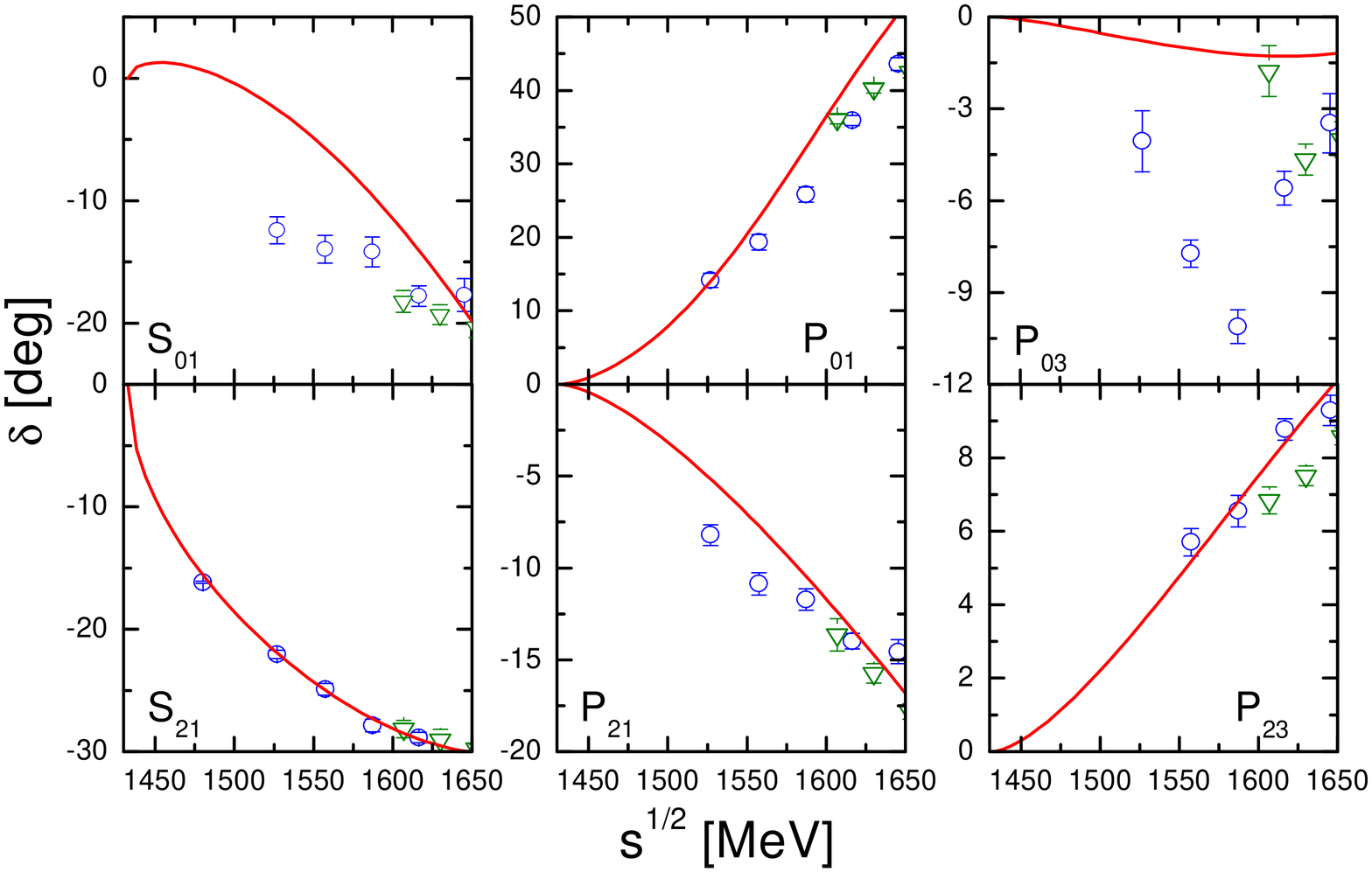,width=.48\textwidth,silent=,clip=}}
\caption{\label{fig:pik}
Left panel:
S- and p-wave pion-nucleon phase shifts. The single energy phase shifts are taken
from \cite{pion-phases}.
Right panel:
S- and p-wave $K^+$-nucleon phase shifts. The solid lines represent the results of the
$\chi$-BS(3) approach. The open circles are from the Hyslop analysis \cite{Hyslop} and the
open triangles from the Hashimoto analysis \cite{Hashimoto}}
\end{figure}

At chiral order $Q^3$ the number of parameters increases
significantly unless further constraints from QCD are imposed.
A systematic expansion of the interaction kernel in powers of $1/N_c$
leads to a much reduced parameter set. For example the $1/N_c$
expansion leads to only four further parameters describing the refined
symmetry-conserving two-body interaction vertices. This is to be compared with
the ten parameters we established to be relevant at order
$Q^3$ if large $N_c$ sum rules are not imposed. Note that at order $Q^3$ there are
no symmetry-breaking 2-body interaction vertices. To that order the only
symmetry-breaking effects result from the refined 3-point vertices. Here a particularly rich picture emerges.
At order $Q^3$ we  established 23 parameters describing symmetry-breaking effects in the 3-point meson-baryon
vertices.  For instance, to that order the baryon-octet states may couple to the pseudo-scalar mesons
also via pseudo-scalar vertices rather than only via the leading axial-vector vertices. Out of those
23 parameters 16 contribute at the same time to matrix elements of the axial-vector current. Thus in order
to control the symmetry breaking effects, it is mandatory to include constraints from the weak decay widths
of the baryon octet states also. A detailed analysis of the 3-point vertices in the $1/N_c$ expansion of QCD
reveals that in fact only ten parameters, rather
than the 23 parameters, are needed at leading order in that expansion. Since
the leading parameters together with the symmetry-breaking parameters describe at the same
time the weak decay widths of the baryon octet and decuplet ground states,
the number of free parameters does not increase significantly at the $Q^3$ level if
the large $N_c$ limit is applied.


In the left panel of Fig.~\ref{fig:pik} we confront the result of our global fit with the empirical $\pi N$ phase shifts.
All s- and p-wave phase shifts are well reproduced up to $\sqrt{s} \simeq 1300$ MeV
with the exception of the $S_{11}$ phase for which our result agrees with the
partial-wave analysis less accurately. We emphasize that one
should not expect quantitative agreement for $\sqrt{s} > m_N+2\,m_\pi \simeq 1215$ MeV where
the inelastic pion production process, not included in this work, starts. The missing higher
order range terms in the $S_{11}$ phase are expected to be induced by additional inelastic
channels or by the nucleon resonances $N(1520)$ and $N(1650)$.
We confirm the findings of \cite{Kaiser,new-muenchen} that the coupled $SU(3)$ channels, if truncated
at the Weinberg-Tomozawa level, predict considerable strength in the $S_{11}$ channel around
$\sqrt{s} \simeq 1500$ MeV where the phase shift shows a resonance-like structure. Note, however
that it is expected that the nucleon resonances $N(1520)$ and $N(1650)$ couple strongly to each
other  and therefore one should not expect a quantitative description of the $S_{11}$
phase too far away from threshold. Similarly we observe considerable strength in the $P_{11}$ channel
leading to a resonance-like structure around $\sqrt{s} \simeq 1500 $ MeV. We interpret this phenomenon
as a precursor effect of the p-wave $N(1440)$ resonance. We stress that our approach differs significantly
from the recent work \cite{new-muenchen} in which the coupled SU(3) channels are applied to pion induced
$\eta$ and kaon production which require much larger energies $\sqrt{s} \simeq m_\eta +m_N \simeq$ 1486 MeV
or $\sqrt{s} \simeq m_K +m_\Sigma \simeq $ 1695 MeV. We believe that such high energies can be
accessed reliably only by including more inelastic channels. It may be worth mentioning that the
inclusion of the inelastic channels as required by the $SU(3)$ symmetry leaves the $\pi N$ phase shifts basically
unchanged for $\sqrt{s}< 1200$ MeV.

In the right panel of  Fig.~\ref{fig:pik} we confront our s- and p-wave $K^+$-nucleon
phase shifts with the most recent analyses by Hyslop et al. \cite{Hyslop} and Hashimoto \cite{Hashimoto}.
We find that our partial-wave phase shifts are reasonably close to the single energy phase shifts of
\cite{Hyslop} and \cite{Hashimoto} except the $P_{03}$ phase for which we obtain much smaller strength.
Note however, that at higher energies we smoothly reach the single energy phase shifts of Hashimoto \cite{Hashimoto}.

\begin{figure}[t]
\parbox{.55\textwidth}{\epsfig{file= 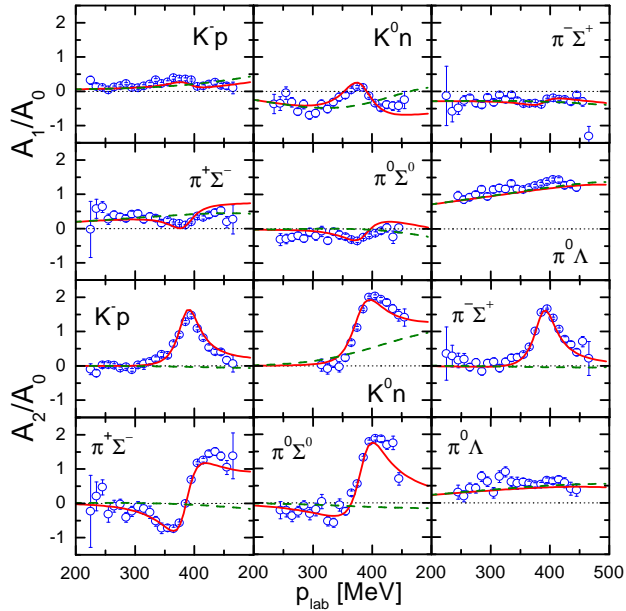,width=.5\textwidth,silent=,clip=}}
\hfill
\parbox{.4\textwidth}{\caption{\label{fig:adep}
Coefficients $A_1$ and $A_2$ for the $K^-p\to \pi^0 \Lambda$,
$K^-p\to \pi^\mp \Sigma^\pm$
and $K^-p\to \pi^0 \Sigma$ differential cross sections, where
$\frac{d\sigma (\sqrt{s}, \cos \theta )}{d\cos \theta }  =
\sum_{n=0}^\infty A_n(\sqrt{s}\,)\,P_n(\cos \theta )
$. The data are
taken from \cite{Adep}. The solid lines are the result of the $\chi$-BS(3) approach
with inclusion of the d-wave resonances. The dashed lines show the effect of switching off d-wave contributions.
}}

\end{figure}

In Fig.~\ref{fig:adep} we compare the empirical ratios $A_1/A_0$ and $A_2/A_0$ of
the inelastic $K^-p$ scattering with the results of
the $\chi$-BS(3) approach. Note that for $p_{\rm lab} < 300$ MeV the empirical ratios with $n\geq 3$ are
compatible with zero within their given errors.  A large $A_1/A_0$ ratio
is found only in the $K^-p\to \pi^0 \Lambda$ channel demonstrating the importance of
p-wave effects in the isospin one channel. The dashed lines of Fig.~\ref{fig:adep}, which are obtained when
switching off d-wave contributions, confirm the importance of this resonance for the angular
distributions in the isospin zero channel.


\end{document}